\documentclass[prd,showpacs,preprint]{revtex4}
\usepackage{amssymb}
\usepackage{amsmath}
\usepackage{graphicx}
\usepackage{multirow}
\usepackage{subfigure}
\usepackage{float}

\begin{document}

\title{Next-to-leading order QCD corrections to the top quark associated with $\gamma$ production via model-independent
flavor-changing neutral-current couplings at hadron colliders}
\author{Yue Zhang}
\author{Bo Hua Li}
\author{Chong Sheng Li}
\email{csli@pku.edu.cn}
\author{Jun Gao}
\author{Hua Xing Zhu}
\affiliation{Department of Physics and State Key Laboratory of
Nuclear Physics and Technology, Peking University, Beijing 100871,
China}

\date{\today}

\pacs{14.65.Ha, 12.38.Bx, 12.60.Cn}

\begin{abstract}
We present the complete next-to-leading order (NLO) QCD corrections
to the top quark associated with $\gamma$ production induced by
model-independent $tq\gamma$ and $tqg$ flavor-changing
neutral-current (FCNC) couplings at hadron colliders, respectively.
We also consider the mixing effects between the $tq\gamma$ and $tqg$
FCNC couplings for this process. Our results show that, for the
$tq\gamma$ couplings, the NLO QCD corrections can enhance the total
cross sections by about $50\%$ and $40\%$ at the Tevatron and LHC,
respectively. Including the contributions from the $tq\gamma$, $tqg$
 FCNC couplings and their mixing effects, the NLO QCD corrections can enhance the
total cross sections by about $50\%$ for the $tu\gamma$ and $tug$
FCNC couplings, and by about the $80\%$ for the $tc\gamma$ and $tcg$
FCNC couplings at the LHC, respectively. Moreover, the NLO
corrections reduce the dependence of the total cross section on the
renormalization and factorization scale significantly. We also
evaluate the NLO corrections for several important kinematic
distributions.
\end{abstract}

\maketitle

\section{Introduction}\label{s1}
Top quark is an excellent probe for the new physics beyond the
standard model (SM), since it is the heaviest particle discovered so
far, with a mass close to the electroweak (EW) symmetry breaking
scale. Direct evidence for new physics at TeV scale may be not easy
to find, while indirect evidence, such as modification of SM
predictions originated from new physics interaction, are important
as well. A good consideration is to investigate single top quark
production process via the anomalous flavor-changing neutral-current
(FCNC) coupling. The FCNC couplings are absent at the tree level,
and occur through loop diagrams within the SM, which are further
suppressed by the Glashow-Iliopoulos-Maiani (GIM)
mechanism~\cite{Glashow:1970gm}. Therefore, within the SM, single
top quark FCNC production is expected to have tiny cross section,
and is probably unmeasurable at the CERN Large Hadron Collider
(LHC). However, the single top quark production induced by the FCNC
coupling can be enhanced significantly in some new physics
models~\cite{delAguila:1998tp, Cheng:1987rs, Li:1993mg,
Davoudiasl:2001uj, HongSheng:2007ve, Wang:1994qd}. Top quark will be
copiously produced at the LHC (about $10^{8}$ per year), even in the
initial low luminosity run ($\sim 10{\rm \ fb}^{-1}$/year)
$8\times10^6$ top quark pairs and $3\times10^6$ single top quarks
will be produced yearly. With such large samples, precise
measurements of single top quark production will be available and
provide a good opportunity to discover the first hint of new physics
by observing the FCNC couplings in the top quark sector.

In this paper, we study the single top quark associated with
$\gamma$ production induced by FCNC couplings in a model-independent
way by using the effective Lagrangian. The relevant effective
lagrangian up to dimension $5$ consists of the following
operators~\cite{Beneke:2000hk}:
\begin{eqnarray}\label{lag}
{\mathcal{L}_{\mathrm{eff}}}
&=&-e\sum_{q=u,c}\frac{\kappa^{\gamma}_{\mathrm{tq}}}{\Lambda}\bar{q}\sigma^{\mu\nu}
(f^{\gamma}_{\mathrm{tq}}+ih^{\gamma}_{\mathrm{tq}}\gamma_5)tA_{\mu\nu}\notag\\
&&-g_s\sum_{q=u,c}\frac{\kappa^g_{\mathrm{tq}}}{\Lambda}\bar{q}\sigma^{\mu\nu}T^a
(f^g_{\mathrm{tq}}+ih^g_{\mathrm{tq}}\gamma_5)tG_{\mu\nu}^a+\mathrm{H.c}.
\end{eqnarray}
where $\Lambda$ is the new physics scale, $A_{\mu\nu}$ and
$G_{\mu\nu}^a$ are the field strength tensors of photon and gluon
respectively, and $T^a$ are the conventional Gell-Mann matrices.
$\kappa^{V}_{tq}$ are real and positive, while $f^V_{\mathrm{tq}}$
and $h^V_{\mathrm{tq}}$ are complex numbers satisfying
$|f^V_{\mathrm{tq}}|^2+|h^V_{\mathrm{tq}}|^2=1$ with $V=\gamma,g$
and $q = u, c$.

Present experimental constraints for the $tq\gamma$ FCNC couplings
come from the non-observation of the decays $t\rightarrow q\gamma$
at Tevatron and the absence of the single top production
$eu\rightarrow et$ at HERA. The CDF collaboration has set $95\%$
confidence level (CL) limits on the branching fractions
$\rm{Br}(t\rightarrow q\gamma)\leq0.032$~\cite{Abe:1997fz}, which
corresponds to $\kappa^{\gamma}_{tq}/\Lambda\leq0.77\rm{TeV^{-1}}$
based on the theoretical predictions of $t\rightarrow q+\gamma$ at
the Next-to-leading order(NLO) level in
QCD~\cite{Zhang:2008yn,Zhang:2010bm}. The ZEUS collaboration also
provides a more stringent constraints, $\kappa^{\gamma}_{tu}<0.174$
at a $95\%$ CL, through the measurements of $ep\rightarrow
etX$~\cite{Chekanov:2003yt} using the NLO
predictions~\cite{Belyaev:2001hf}, which can be transferred to
$\kappa^{\gamma}_{tu}/\Lambda<0.33\rm{TeV^{-1}}$. Recently, the most
stringent experimental constraints for the $tqg$ FCNC couplings are
$\kappa^g_{tu}/\Lambda\leq0.013\ {\rm TeV^{-1}}$ and
$\kappa^g_{tc}/\Lambda\leq 0.057\ {\rm TeV^{-1}}$ given by the D0
Collaboration~\cite{Abazov:2010qk}, and
$\kappa^g_{tu}/\Lambda\leq0.018\ {\rm TeV^{-1}}$ and
$\kappa^g_{tc}/\Lambda\leq 0.069\ {\rm TeV^{-1}}$ given by the CDF
Collaboration~\cite{Aaltonen:2008qr}, based on the measurements of
the FCNC single top production using the theoretical predictions,
including the NLO QCD corrections~\cite{Liu:2005dp,Gao:2009rf} and
resummation effects~\cite{Yang:2006gs}, respectively.

The observation of $qg\rightarrow\gamma t$ process is a clear signal
of top quark FCNC interactions, which can be induced by $tq\gamma$
and $tqg$ couplings, since there is no irreducible backgrounds of
this process in the SM. There are already several
literatures~\cite{AguilarSaavedra:2004wm} discussing this process
using effective Lagrangian Eq.~(\ref{lag}). However they were either
based on the LO calculations, or the NLO QCD effects are not
completely calculated. So it is necessary to present a complete NLO
corrections to the above process, which is not only mandatory for
matching the expected experimental accuracy at hadron colliders, but
is also important for a consistent treatment of both the top quark
production and decay via the FCNC couplings by experiments.

In this paper, we present the complete NLO QCD corrections to the
top quark associated with $\gamma$ production via $tq\gamma$ and
$tqg$ FCNC couplings with their mixing effects at hadron colliders.

The arrangement of this paper is as follows. In Sec.~\ref{s2}, we
present the LO results for the top quark associated with $\gamma $
production induced by the $tq\gamma$ FCNC couplings. In
Sec.~\ref{s3}, we show the details of the corresponding NLO
calculations. Sec.~\ref{s4} contains the analysis of
$qg\rightarrow\gamma t$ process induced by the the $tq\gamma$, $tqg$
FCNC couplings and the mixing effects. We present the numerical
results in Sec.~\ref{s5}. Finally, we give our conclusion in
Sec.~\ref{s6}.

\section{Leading order results}\label{s2}
\begin{figure}[ht]
\begin{center}
\scalebox{1.0}{\includegraphics*[width=0.6\linewidth]{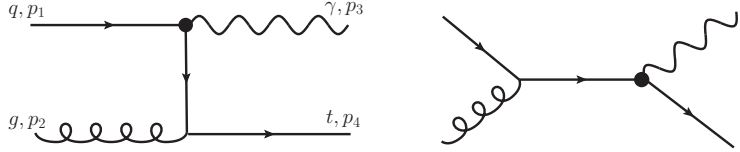} }
\caption[]{\label{f1}The LO Feynman diagrams for top quark
associated with $\gamma$ production via the $tq\gamma$ FCNC
couplings.}
\end{center}
\end{figure}

At hardron colliders, there is only one process $qg\rightarrow\gamma
t$ with $q=c,u$ that contributes to the $t\gamma$ associated
production at the LO via the electroweak FCNC couplings,
$\kappa^{\gamma}_{tq}$. The corresponding Feynman diagrams are shown
in Fig.~\ref{f1}.

From the effective operator in Eq.~(\ref{lag}), we obtain the
following LO squared amplitudes for above process in four
dimensions,
\begin{eqnarray}\label{eq1}
\overline{|M^B|^2}(s,t,u)&=&
\frac{64\pi^{2}\alpha_{s}\alpha}{3s(m^{2}-t)^{2}}(\frac{
\kappa^{\gamma}_{tq}}{\Lambda})^{2}
\Big[m^{8}-(2s+t)m^{6}+(s^{2}+4st+t^{2})m^{4}\nonumber \\
&&-(3s^{2}+6st+t^{2})tm^{2}+2st^{2}(s+t)\Big],
\end{eqnarray}
where $m$ is the top quark mass, the colors and spins of the
outgoing particles have been summed over, and the colors and spins
of the incoming ones have been averaged over, $s$, $t$, and $u$ are
Mandelstam variables, which are defined as
\begin{equation}
s=(p_1+p_2)^2,\ \ t=(p_1-p_3)^2,\ \ u=(p_1-p_4)^2.
\end{equation}
After the phase space integration, the LO partonic cross
sections
are given by
\begin{equation}
\hat \sigma^B_{ab}=\frac{1}{2\hat s}\int d\Gamma
\overline{|M^B|^2}_{ab}.
\end{equation}
The LO total cross section at hadron colliders is obtained
by
convoluting the partonic cross section with the parton
distribution
functions (PDFs) $G_{i/P}$ for the proton (antiproton):
\begin{equation}
\sigma^B=\sum_{ab}\int dx_1
dx_2\left[G_{a/P_1}(x_1,\mu_f)G_{b/P_2}(x_2,\mu_f)\hat
\sigma^B_{ab}\right],
\end{equation}
where $\mu_f$ is the factorization scale.

\section{Next-to-leading order QCD corrections}\label{s3}
\begin{figure}[ht]
\begin{center}
\scalebox{1.0}{\includegraphics*[width=0.9\linewidth]{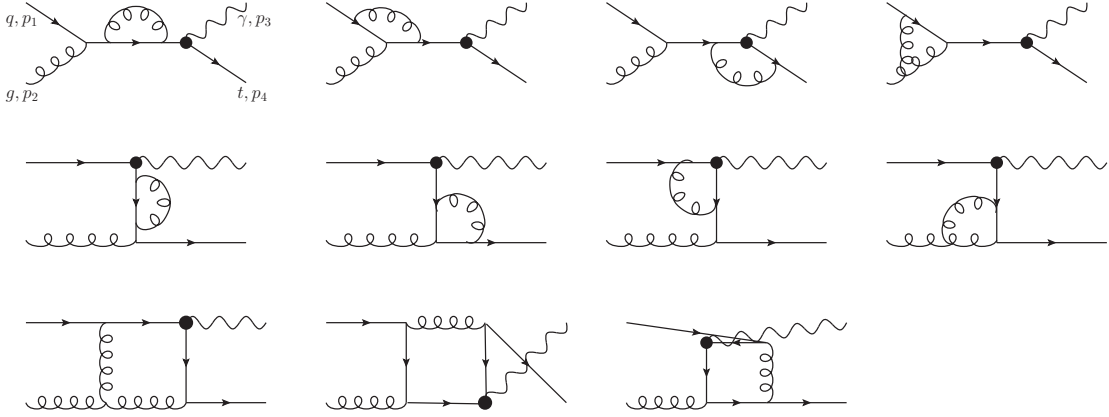}}
\caption[]{\label{f2}One-loop Feynman diagrams for the process
$qg\to \gamma t$ induced by the $tq\gamma$ FCNC couplings.}
\end{center}
\end{figure}

\begin{figure}[ht]
\begin{center}
\scalebox{1.0}{\includegraphics*[width=0.9\linewidth]{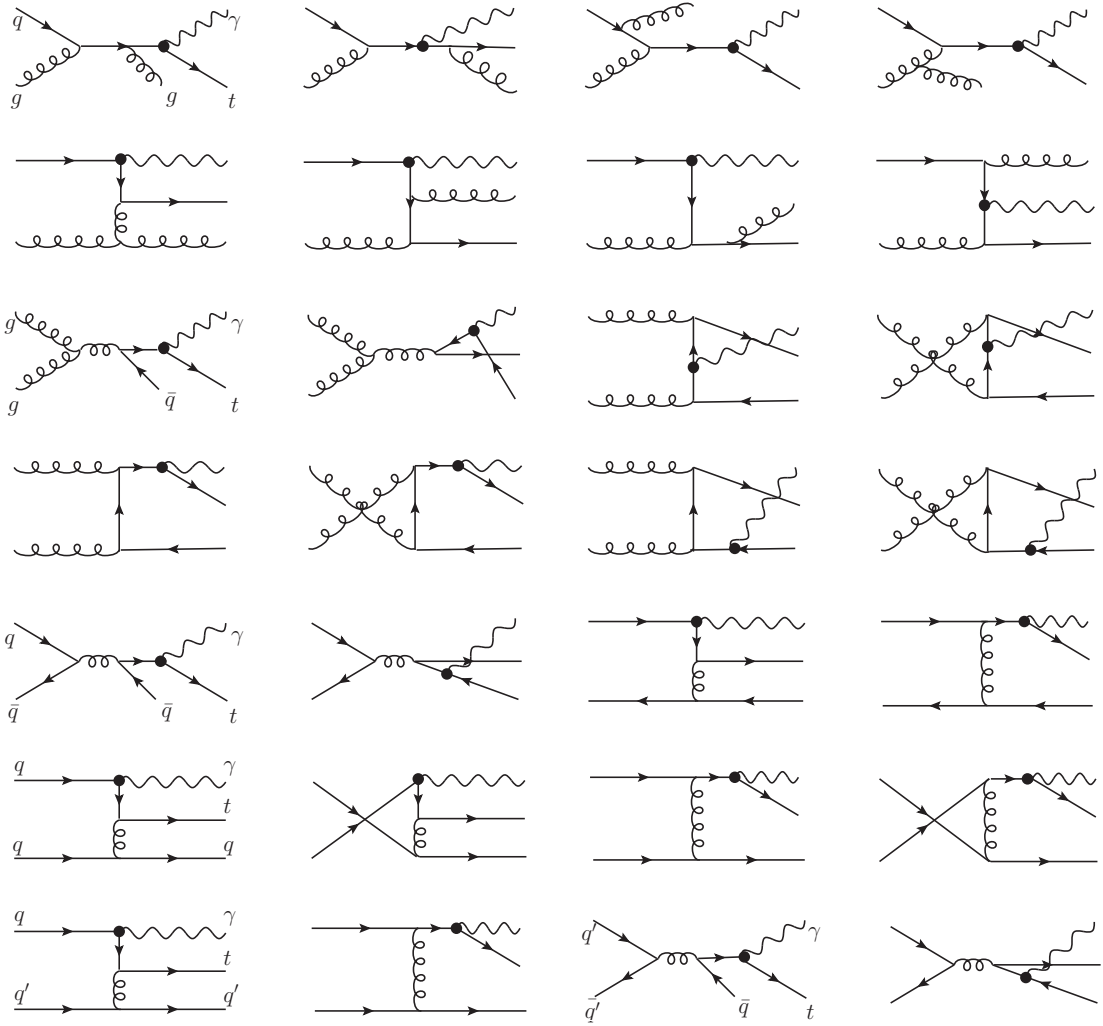} }
\caption[]{\label{f3}Feynman diagrams of the real corrections
induced by the $tq\gamma$ FCNC couplings.}
\end{center}
\end{figure}

In this section, we present our calculations for the NLO QCD
correcions to the top quark associated with $\gamma$ production via
the electroweak FCNC couplings. At the NLO, we need to include
contributions from both the virtual corrections (Fig.~\ref{f2}) and
the real corrections (Fig.~\ref{f3}). We use the dimensional
regularization scheme (with naive $\gamma_{5}$) in $n=4-2\epsilon$
dimensions to regularize both ultraviolet (UV) and infrared (IR)
divergence. Moreover, for the real corrections, we used the dipole
subtraction method with massive
partons~\cite{Catani:1996vz,Catani:2002hc} to separate the IR
divergence.

All the UV divergence appearing in the loop diagrams are canceled by
introducing counterterms for the wave functions and mass of the
external fields ($\delta Z_2^{(g)},\delta Z_2^{(q)},\delta
Z_2^{(t)},\delta m$), and the coupling constants ($\delta
Z_{g_s},\delta Z_{\kappa^\gamma_{tq}/\Lambda}$). We define these
counterterms according to the same procedures adopted in
Ref.\cite{Liu:2005dp}:
\begin{eqnarray}\label{rencons}
\delta Z_2^{(g)}&=&-\frac{\alpha_s}{2\pi}C_{\epsilon}\left(
\frac{N_f}{3}-\frac{5}{2}\right)\left(
\frac{1}{\epsilon_{\mathrm{UV}}}
-\frac{1}{\epsilon_{\mathrm{IR}}}\right)-\frac{\alpha_s}{
6\pi}C_{\epsilon}
\frac{1}{\epsilon_{\mathrm{UV}}},\notag\\
\delta Z_2^{(q)}&=&-\frac{\alpha_s}{3\pi}C_{\epsilon}\left(
\frac{1}{\epsilon_{\mathrm{UV}}}
-\frac{1}{\epsilon_{\mathrm{IR}}}\right),\notag\\
\delta Z_2^{(t)}&=&-\frac{\alpha_s}{3\pi}C_{\epsilon}\left(
\frac{1}{\epsilon_{\mathrm{UV}}}+
\frac{2}{\epsilon_{\mathrm{IR}}}+4\right),\notag\\
\delta
Z_{g_s}&=&\frac{\alpha_s}{4\pi}\Gamma(1+\epsilon)(4\pi)^{
\epsilon}
\left(\frac{N_f}{3}-\frac{11}{2}\right)
\frac{1}{\epsilon_{\mathrm{UV}}}+\frac{\alpha_s}{12\pi}
C_{\epsilon}\frac{1}{\epsilon_{\mathrm{UV}}},
\end{eqnarray}
where $C_{\epsilon}=\Gamma(1+\epsilon)[(4\pi\mu_r^2)/m^2]
^{\epsilon}$ and $n_f=5$. For $\delta m$, we use the on-shell
subtraction:
\begin{equation}\label{renmt}
\frac{\delta
m}{m}=-\frac{\alpha_s}{3\pi}C_{\epsilon}\left(\frac
{3}{\epsilon_{UV}}+4\right),
\end{equation}
And, we adopt the $\rm\overline{MS}$ scheme for the renormalization
constants of the electroweak FCNC couplings $\delta
Z_{\kappa^{\gamma}_{tq}/\Lambda}$, and adjust it to cancel the
remaining UV divergence exactly:
\begin{equation}\label{renkgamma}
\delta
Z_{\kappa^{\gamma}_{tq}/\Lambda}=\frac{\alpha_s}{3\pi}
\Gamma(1+\epsilon)(4\pi)
^{\epsilon}\frac{1}{\epsilon_{UV}},
\end{equation}

Here we first consider the electroweak FCNC couplings, the running
of the couplings are given by~\cite{Zhang:2010bm}
\begin{equation}
\kappa^{\gamma}(\mu)=\kappa^{\gamma}(\mu')
\eta^{\frac{4}{3\beta_0}},
\end{equation}
where $\eta=\alpha_s(\mu')/\alpha_s(\mu)$ and $\beta_0$ is the
1-loop QCD $\beta$-function given by $11-\frac{2}{3}n_f$ with $n_f$
active flavors between the two scales $\mu$ and $\mu^{\prime}$.

The squared amplitudes of the virtual corrections are
\begin{equation}\label{eq5}
\overline{|M|^2}_{1-loop}=\sum_i2Re\overline{(M^{loop,i}M^{
B*})}
+2Re\overline{(M^{con}M^{B*})},
\end{equation}
where $M^{loop,i}$ denote the amplitudes for the $i$-th loop diagram
in Fig.~\ref{f2}, and $M^{con}$ are the corresponding counterterms.
All the UV divergence in Eq.~(\ref{eq5}) have been cancelled as they
must, but the IR divergent pieces are still present. Because of the
limited space, we do not show the lengthy explicit expressions of
the virtual corrections here. The IR divergence of the virtual
corrections can be factorized as
\begin{eqnarray}\label{eq2}
\overline{|M|^2}_{one-loop,IR}&=&-\frac{\alpha_s}{12\pi}D_{
\epsilon}
\bigg\{\frac{13}{\epsilon_{IR}^2} +
\left[9\ln(\frac{s}{m^{2}})+9\ln(\frac{m^{2}-t}{m^{2}}
)-\ln(\frac{s+t}{m^{2}})+\frac{43}{2}\right]
\frac{1}{\epsilon_{IR}}\bigg\} \nonumber \\
&&\times \overline{|M^B|^2},
\end{eqnarray}
where $D_{\epsilon}=[(4\pi\mu_r^2)/m^{2}]^
{\epsilon}/\Gamma(1-\epsilon)$, and $\overline{|M^B|^2}$ are
the
squared Born amplitudes given in Eq.~(\ref{eq1}).

At the NLO the real corrections consist of the radiations of an
additional gluon or massless (anti)quark in the final states,
including the subprocesses
\begin{eqnarray}
&&q\ g\ \rightarrow\gamma\ t\ g,\ g\ g\ \rightarrow \gamma\
t\
\bar{q},\ q\ \bar{q}\ \rightarrow \gamma\ t\ \bar{q},
\nonumber \\
&&q\ q\ \rightarrow \gamma\ t\ q,\ q\ q'\ \rightarrow
\gamma\ t\
q',\ q'\ \bar{q'}\ \rightarrow \gamma\ t\ \bar{q},
\end{eqnarray}
here $q$ denotes $u$ quark for the $\kappa^{\gamma}_{tu}$ coupling
and $c$ quark for the $\kappa^{\gamma}_{tc}$ coupling, while $q'$
denotes massless (anti)quark other than $q$. It should be noted that
in our NLO calculations we do not include the contributions from the
SM on-shell production of the top pair with subsequent rare decay of
one top quark, $pp(\bar p) \to t\bar t\to \gamma +t +\bar q$, which
provide the same signature as the top quark associated with $\gamma$
production via the FCNC couplings and can be calculated separately.

Before performing the numerical calculations, we need to extract the
IR divergences in the real corrections. In the dipole formalism this
is done by subtracting some dipole terms from the real corrections
to cancel the singularities exactly, such that the real corrections
become integrable in four dimensions. These dipole subtraction terms
are analytically integrable in $n$ dimensions over one-parton
subspaces, which give $\epsilon$ poles that represent the soft and
collinear divergences. Then we can add them to the virtual
corrections to cancel the $\epsilon$ poles, and ensure the virtual
corrections are also integrable in four dimensions. This whole
procedure can be illustrated by the formula~\cite{Catani:2002hc}:
\begin{equation}
\hat{\sigma}^{NLO}=\int_{m+1}\left[\left(d\hat{\sigma}^R
\right)_{\epsilon=0}-\left(d\hat{\sigma}^A\right)_{
\epsilon=0}\right]
+\int_m\left[d\hat{\sigma}^V+\int_1d\hat{\sigma}^A\right]_{
\epsilon=0},
\end{equation}
where $m$ is the number of final state particles at the LO,
and
$d\hat{\sigma}^A$ is a sum of the dipole terms. Besides, at
hadron
colliders, we have to include the well-known collinear
subtraction
counterterms in order to cancel the collinear divergences
arising
from the splitting processes of the initial state massless
partons.
Here we use the $\overline{\rm MS}$ scheme and the
corresponding NLO
PDFs.

For the process with two initial state hadrons, the dipole
terms can
be classified into four groups, the final-state emitter and
final-state spectator type,
\begin{eqnarray}
&&\mathcal {D}_{ij,k}(p_1,...,p_{m+1})=\nonumber\\
&&\qquad -\frac{1}{(p_i+p_j)^2-m_{ij}^2}{\
_{m}}\langle...,\widetilde{ij},...,\widetilde
k,...|\frac{\textbf{T}_k\cdot\textbf{T}_{ij}}{\textbf{T}_{ij
}^2}
{\textbf{V}_{ij,k}}|..., \widetilde{ij},...,\widetilde
k,...\rangle_{m},
\end{eqnarray}
the final-state emitter and initial-state spectator type,
\begin{eqnarray}
&&\mathcal {D}_{ij}^a(p_1,...,p_{m+1};p_a,...)=\nonumber\\
&&\qquad
-\frac{1}{(p_i+p_j)^2-m_{ij}^2}\frac{1}{x_{ij,a}} {\
_{m,a}}\langle...,\widetilde{ij},...;\widetilde
a,...|\frac{\textbf{T}_a\cdot\textbf{T}_{ij}}{\textbf{T}_{ij
}^2}
{\textbf{V}_{ij}^a}|..., \widetilde{ij},...;\widetilde
a,...\rangle_{m,a},
\end{eqnarray}
the initial-state emitter and final-state spectator type,
\begin{eqnarray}
&&\mathcal {D}_{j}^{ai}(p_1,...,p_{m+1};p_a,...)=\nonumber\\
&&\qquad -\frac{1}{2p_a p_i}\frac{1}{x_{ij,a}} {\
_{m,\widetilde{ai}}}\langle...,\widetilde{j},...;\widetilde
{ai},...|\frac{\textbf{T}_j\cdot\textbf{T}_{ai}}{\textbf{T}_
{ai}^2}
{\textbf{V}_{j}^{ai}}|..., \widetilde{j},...;\widetilde
{ai},...\rangle_{m,\widetilde{ai}},
\end{eqnarray}
and the initial-state emitter and initial-state spectator
type,
\begin{eqnarray}
&&\mathcal {D}^{ai,b}(p_1,...,p_{m+1};p_a,p_b)=\nonumber\\
&&\qquad -\frac{1}{2p_a p_i}\frac{1}{x_{i,ab}} {\
_{m,\widetilde{ai}}}\langle...;\widetilde
{ai},b|\frac{\textbf{T}_b\cdot\textbf{T}_{ai}}{\textbf{T}_{
ai}^2}
{\textbf{V}^{ai,b}}|...;\widetilde
{ai},b\rangle_{m,\widetilde{ai}},
\end{eqnarray}
where $a,b$ and $i,j,...$ are the initial and final state
partons,
and $\textbf{T}$ and $\textbf{V}$ are the color charge
operators and
dipole functions acting on the LO amplitudes, respectively.
The
explicit expressions for $x_{i,ab}$, $x_{ij,a}$ and
$\textbf{V}$ can
be found in Ref.~\cite{Catani:2002hc}. The integrated dipole
functions together with the collinear counterterms can be
written in
the following factorized form
\begin{eqnarray}\label{eq4}
\sim &&\int d\Phi^{(m)}(p_a,p_b) \
_{m,ab}\langle...;p_a,p_b|\textbf{I}_{m+a+b}(\epsilon)|
...;p_a,p_b\rangle_{m,ab}\nonumber\\
&&+\sum_{a'}\int_0^1 dx\int d\Phi^{(m)}(xp_a,p_b)
_{m,a'b}\langle
...;xp_a,p_b|\textbf{P}_{m+b}^{a,a'}
(x)+\textbf{K}_{m+b}^{a,a'}(x)|...;xp_a,p_b\rangle_{m,a'b}
\nonumber\\
&&\qquad \qquad \qquad +(a\leftrightarrow b),
\end{eqnarray}
where $x$ is the momentum fraction of the splitting parton,
$d\Phi^{(m)}$ contains all the factors apart from the
squared
amplitudes, $\textbf{I}$, $\textbf{P}$, and $\textbf{K}$ are
insertion operators defined in~\cite{Catani:2002hc}.

The operators $\textbf{P}$ and $\textbf{K}$ provide finite
contributions to the NLO corrections, and only the operator
$\textbf{I}$ contains the IR divergences
\begin{eqnarray}\label{eq3}
\textbf{I}|_{IR}&=&-\frac{\alpha_s}{2\pi}\frac{(4\pi)^{
\epsilon }}{\Gamma(1-\epsilon)}\bigg\{\sum_j\sum_{k\neq
j}\textbf{T}_j\cdot
\textbf{T}_k\bigg[\left(\frac{\mu_r^2}{s_{jk}}\right)^{
\epsilon}
\mathcal{V}(s_{jk},m_j,m_k;\epsilon_{IR})
+\frac{1}{\textbf{T}_j^2}\Gamma_j(m_j,\epsilon_{IR}
)\bigg]\nonumber\\
&& +\sum_j\textbf{T}_j\cdot\textbf{T}_a\bigg[2\left(
\frac{\mu_r^2}{s_{ja}}\right)^{\epsilon}\mathcal{V}(s_{ja},
m_j,0;\epsilon_{IR})
+\frac{1}{\textbf{T}_j^2}\Gamma_j(m_j,\epsilon_{IR})
+\frac{1}{\textbf{T}_a^2}\frac{\gamma_a}{\epsilon_{IR}}
\bigg ]
\nonumber\\&& +\textbf{T}_a\cdot\textbf{T}_b\bigg[\left(
\frac{\mu_r^2}{s_{ab}}\right)^{\epsilon}\left(\frac{1}
{\epsilon_{IR}^2}+\frac{1}{\textbf{T}_a^2}\frac{
\gamma_a }
{\epsilon_{IR}}\right)\bigg]+(a\leftrightarrow b)\bigg\},
\end{eqnarray}
with
\begin{eqnarray}
\mathcal{V}(s_{jk},m_j,m_k;\epsilon_{IR})&=&\frac{1}{v_
{jk}}\left(\frac{Q_{jk}^2}{s_{jk}}\right)^{\epsilon} \times
\left(
1-\frac{1}{2}\rho_j^{-2\epsilon}-\frac{1}{2}\rho_k^{
-2\epsilon}\right)
\frac{1}{\epsilon_{IR}^2},\nonumber\\
\Gamma_j(0,\epsilon_{IR})&=&\frac{\gamma_j}{\epsilon_{IR}},
\quad
\Gamma_j(m_j\neq 0,\epsilon_{IR})=\frac{C_F}{\epsilon_{IR}},
\end{eqnarray}
where $C_F=4/3$, $\gamma_q=2$, and $\gamma_g=11/2-n_f/3$.
And
$s_{jk}$, $Q_{jk}^2$, $v_{jk}$, and $\rho_n$ are kinematic
variables
defined as follows
\begin{eqnarray}
s_{jk}&=&2p_jp_k,\quad Q_{jk}^2=s_{jk}+m_j^2+m_k^2,\quad
v_{jk}=\sqrt{1-\frac{m_j^2m_k^2}{(p_jp_k)^2}}, \nonumber\\
\rho_n&=&\sqrt{\frac{1-v_{jk}+2m_n^2/(Q_{jk}^2-m_j^2-m_k^2)}
{1+v_{jk}+2m_n^2/(Q_{jk}^2-m_j^2-m_k^2)}}\quad (n=j,k).
\end{eqnarray}
When inserting Eq.~(\ref{eq3}) into the LO amplitudes as
shown in
Eq.~(\ref{eq4}), we can see that the IR divergences can be
written
as combinations of the LO color correlated squared
amplitudes and
all the IR divergences from the virtual corrections in
Eq.~(\ref{eq2}) are canceled exactly, as we expected.

In order to check our result, we have also performed the calculation
with the two-cutoff method~\cite{Harris:2001sx}. We find that the
numerical results are in good agreement.

\section{Contributions From The Electroweak and Strong FCNC Couplings with Mixing
Effects}\label{s4} In previous sections, we only consider the
contributions from the electroweak FCNC couplings,
$\kappa^{\gamma}_{tq}$. However, for the top quark associated
$\gamma$ production process, $qg\rightarrow \gamma t$, there are
additional contributions from the strong FCNC couplings,
$\kappa^{g}_{tq}$, and the mixing effects between these two
operators. Because the magnitudes of the coefficients
$\kappa^{V}_{tq}$ ($V=\gamma,g$) depend on the underlying new
physics, these operator mixing contributions may be significant in
certain models. Since the $\mathcal {O}(\alpha_{s})$ corrections to
the process $qg\rightarrow \gamma t$ induced by $\kappa^{g}_{tq}$
are similar to the ones induced by $\kappa^{\gamma}_{tq}$, we don't
show its analytical results, and only present the combination of the
contributions from the $tq\gamma$, $tqg$ and their mixing effects in
this section.

In case that both of the $\kappa^{g}_{tq}$ and
$\kappa^{\gamma}_{tq}$ are at the same order, the terms proportional
to $(\kappa^{g}_{tq})^{2}$, $\kappa^{g}\kappa^{\gamma}$ and
$(\kappa^{\gamma}_{tq})^{2}$ can contribute to $qg\rightarrow \gamma
t$ with the same significance. The mixing terms, which are
proportional to $\kappa^{g}\kappa^{\gamma}$, could appear in both
the LO and NLO.

\begin{figure}[ht]
\begin{center}
\scalebox{1.0}{\includegraphics*[width=0.6\linewidth]{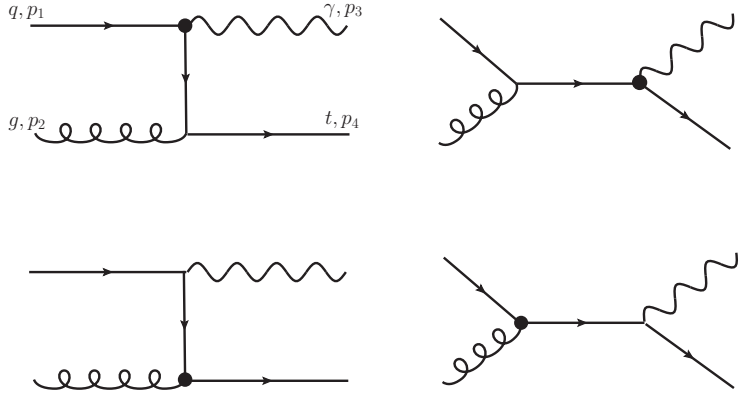} }
\caption[]{\label{f4}The LO Feynman diagrams for top quark
production associated with $\gamma$ via the $tq\gamma$ and $tqg$
FCNC couplings.}
\end{center}
\end{figure}

At the LO, there are four Feynman diagrams for this process, as
shown in Fig.~\ref{f4}. Two of them are the same as Fig.~\ref{f1},
while the others are induced by $\kappa^{g}_{qt}$.

The complete LO squared amplitudes for this process in four
dimensions are
\begin{eqnarray}\label{eq6}
\overline{|M^B|^2_{tot}}(s,t,u)&=&
\frac{16\pi^{2}\alpha_{s}\alpha}{3}\bigg\{\frac{4}{s(m^{2}
-t)^{2}}(\frac{\kappa^{\gamma}_{tq}}{\Lambda})^{2}
\Big[m^{8}-(2s+t)m^{6}+(s^{2}+4st+t^{2})m^{4}\nonumber \\
&&-(3s^{2}+6st+t^{2})tm^{2}+2st^{2}(s+t)\Big] \nonumber
\\&&+\frac{16}{9(m^{2}-s)^{2}t}(\frac{\kappa^{g}_{tq}}{
\Lambda})^{2}\Big[m^{8}-(s+2t)m^{6}+(s^{2}+4st+t^{2})m^{4}\nonumber \\
&&-(s^{2}+6st+3t^{2})sm^{2}+2s^{2}t(s+t)\Big] \nonumber
\\&&-\frac{16}{3(m^{2}-s)(m^{2}-t)}(\frac{\kappa^{g}_{tq}
\kappa^{\gamma}_{tq}}{\Lambda^{2}})
\textrm{Re}(f_{tq}^{\gamma\ast}f_{tq}^{g}+h_{tq}^{\gamma\ast
}h_{tq}^{g})
\Big[3m^{6}-4(s+t)m^{4}\nonumber \\
&&+(s^{2}+3st+t^{2})m^{2}-st(s+t)\Big]\bigg\},
\end{eqnarray}

\begin{figure}[ht]
\begin{center}
\scalebox{1.0}{\includegraphics*[width=0.9\linewidth]{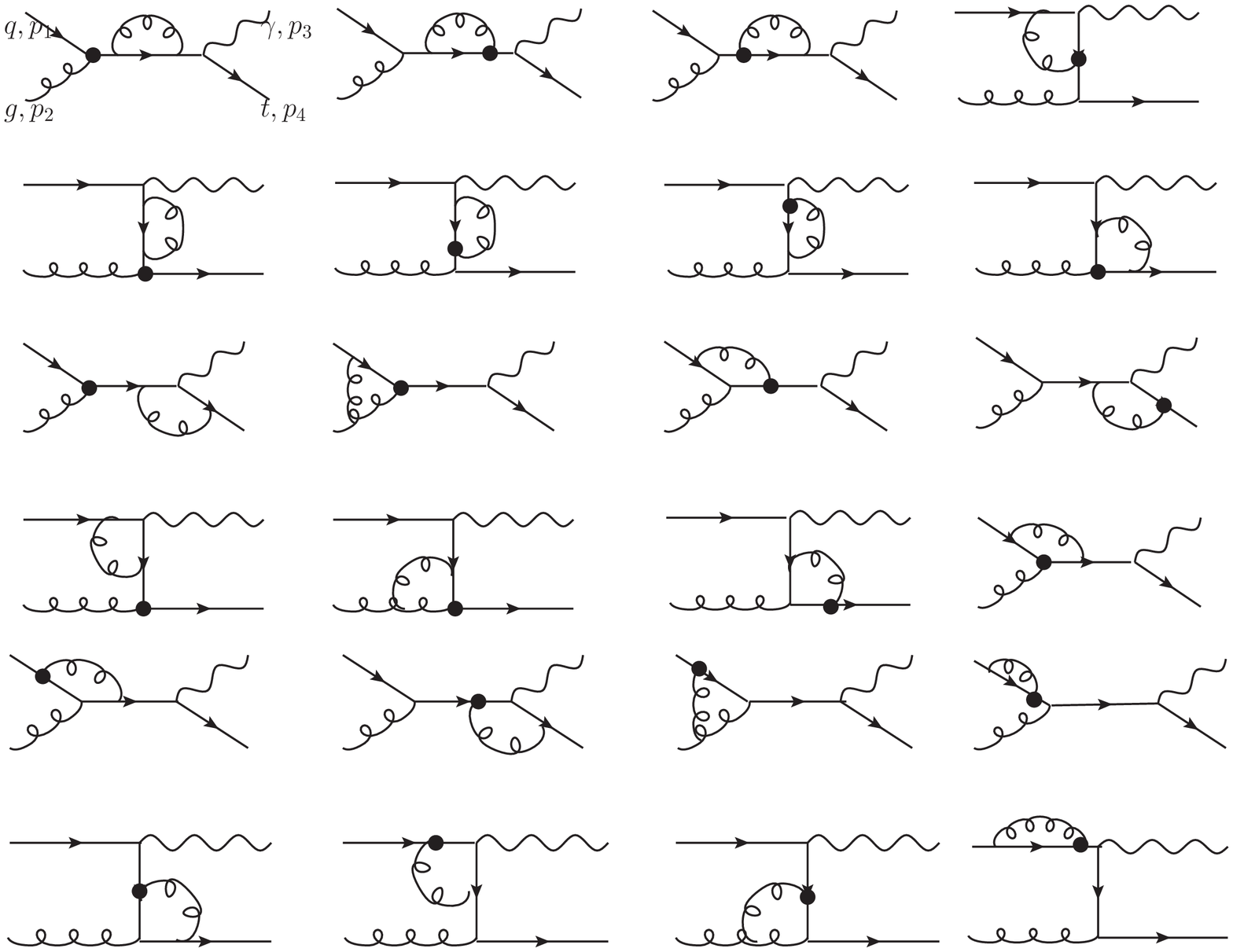}}
\caption[]{\label{fvirtual2}One-loop Feynman diagrams induced by the
$tqg$ FCNC couplings, part I.}
\end{center}
\end{figure}

\begin{figure}[ht]
\begin{center}
\scalebox{1.0}{\includegraphics*[width=0.9\linewidth]{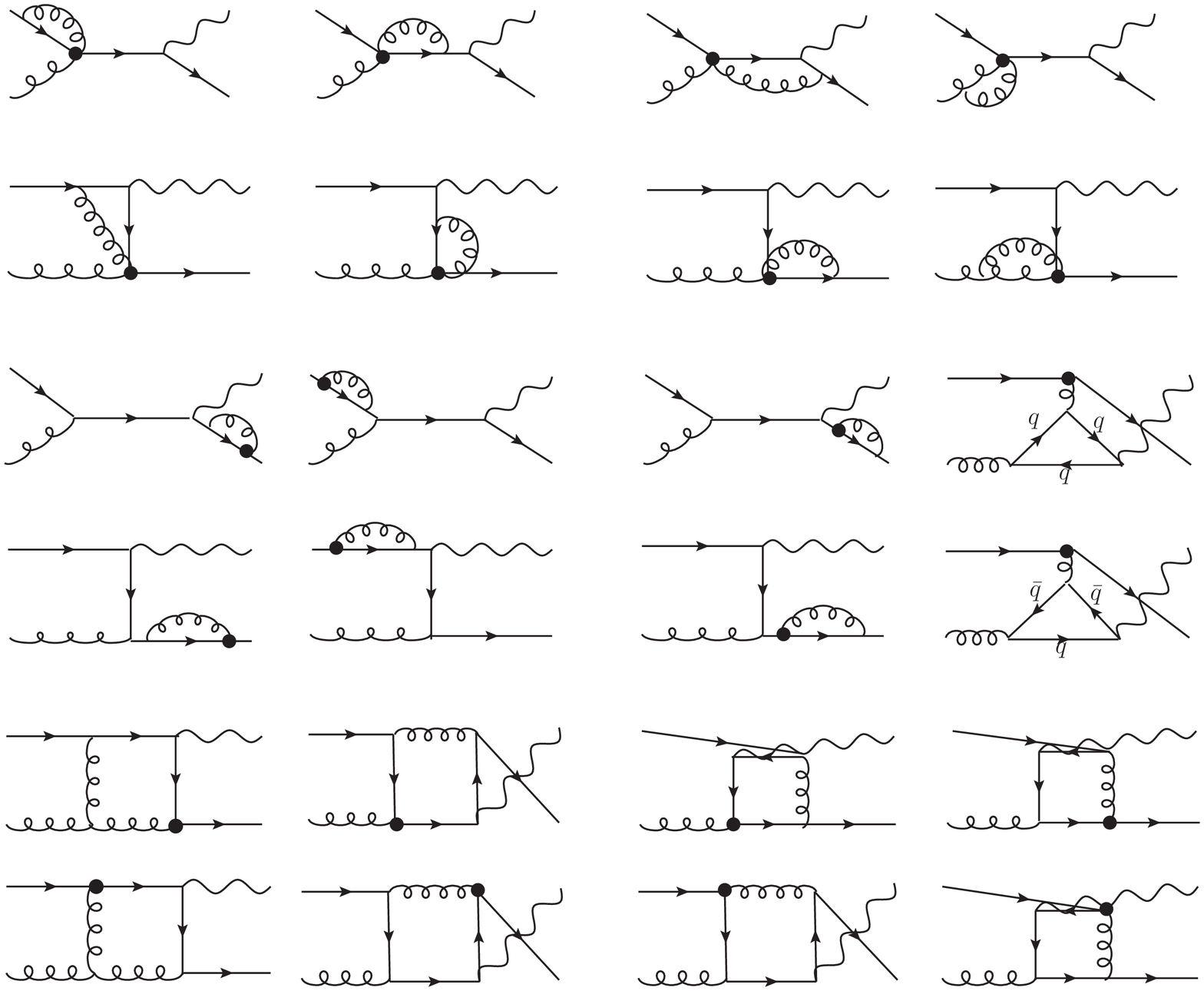}}
\caption[]{\label{fvirtual3}One-loop Feynman diagrams induced by the
$tqg$ FCNC couplings, part II.}
\end{center}
\end{figure}

\begin{figure}[ht]
\begin{center}
\scalebox{1.0}{\includegraphics*[width=0.9\linewidth]{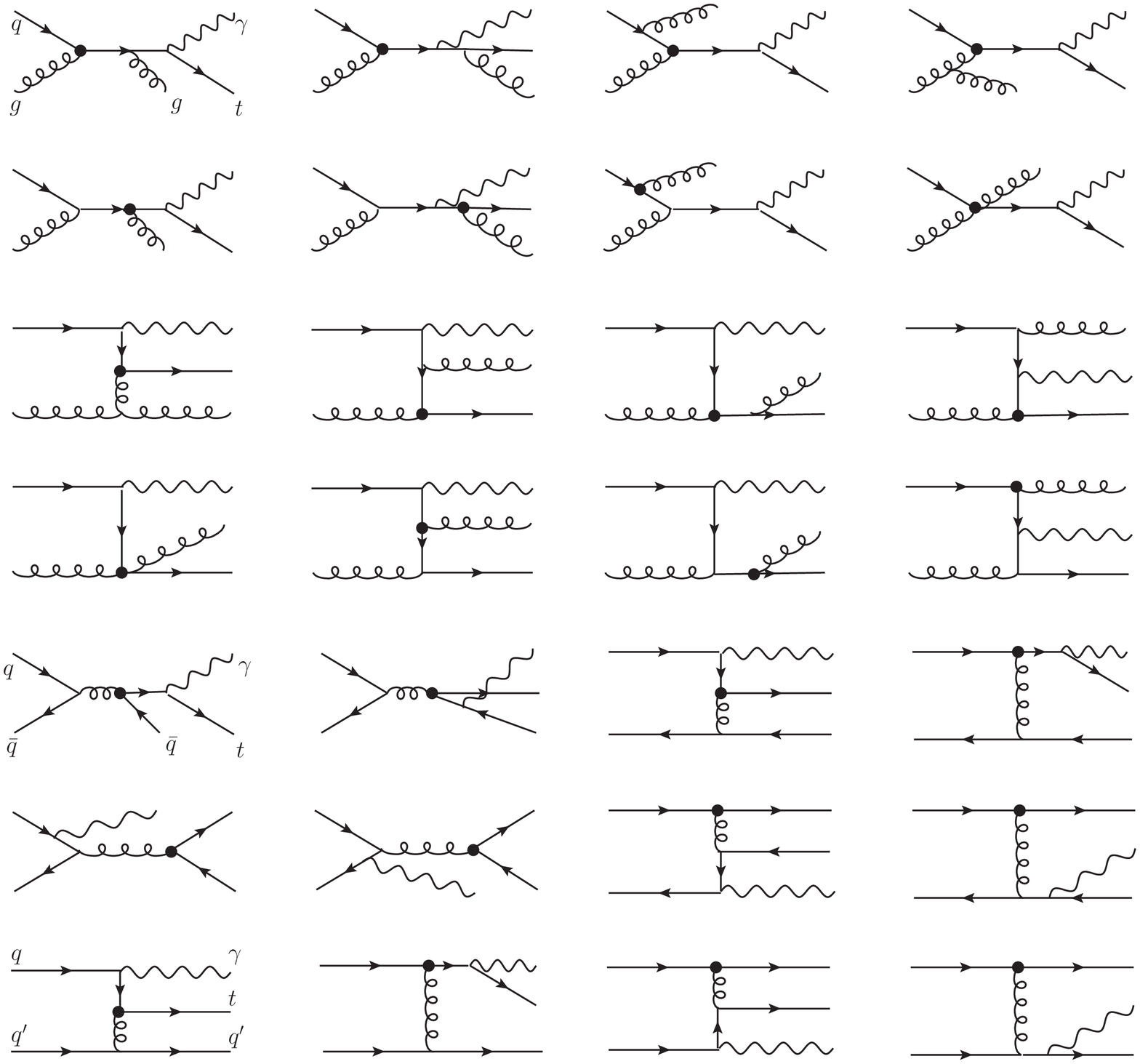}}
\caption[]{\label{freal2}Feynman diagrams of the real corrections
induced by the $tqg$ FCNC couplings, part I.}
\end{center}
\end{figure}

\begin{figure}[ht]
\begin{center}
\scalebox{1.0}{\includegraphics*[width=0.9\linewidth]{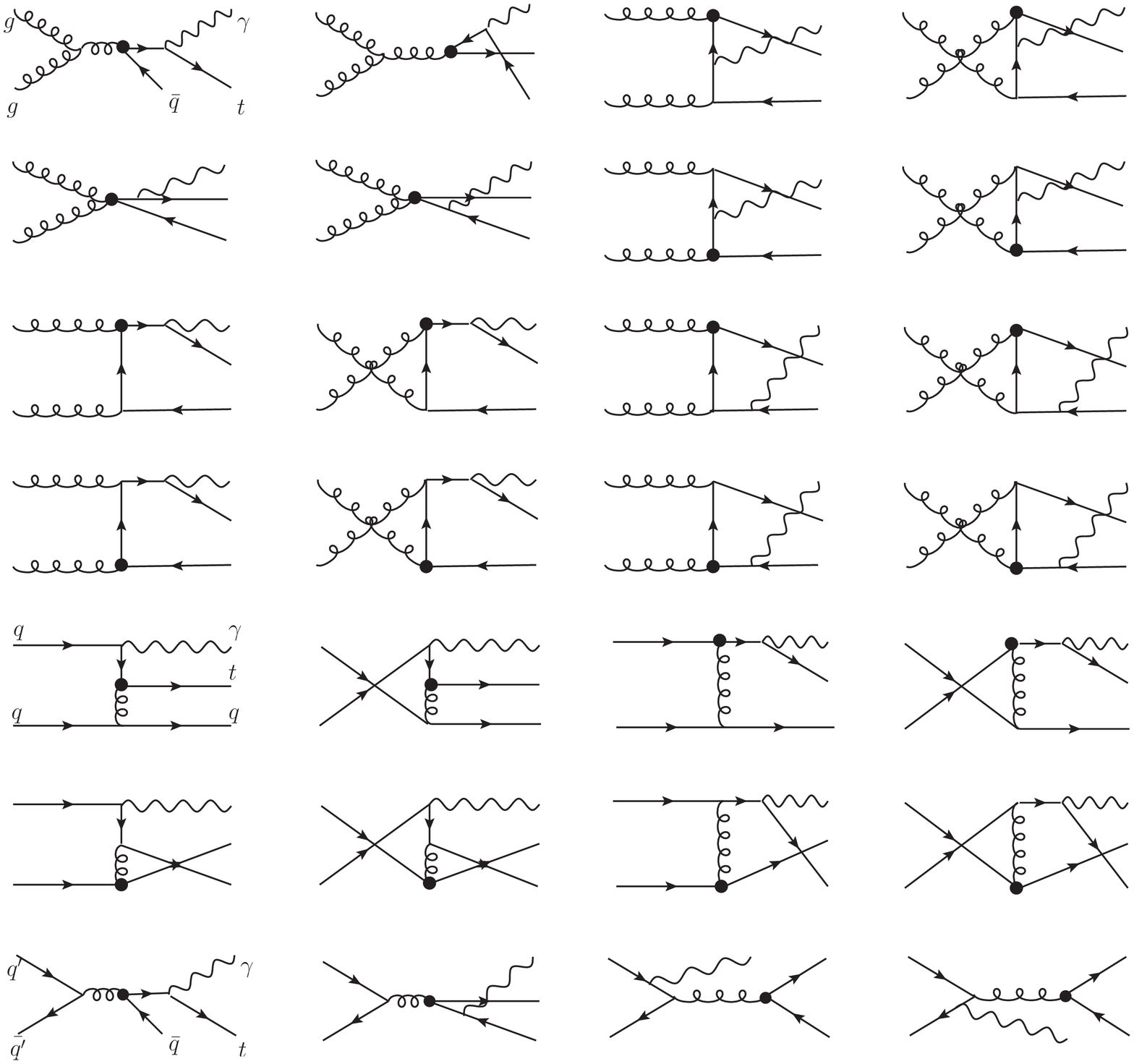}}
\caption[]{\label{freal3}Feynman diagrams of the real corrections
induced by the $tqg$ FCNC couplings, part II.}
\end{center}
\end{figure}

At the NLO, we need to include both the virtual corrections
(Fig.~\ref{f2},~\ref{fvirtual2} and~\ref{fvirtual3}) and the real
corrections (Fig.~\ref{f3},~\ref{freal2} and~\ref{freal3}). The
relevant renormalization constants are the same as ones in
Eqs.~(\ref{rencons}),~(\ref{renmt}) and~(\ref{renkgamma}), except
that we introduce additional renormalization constants. We adopt the
definition in Ref.~\cite{Zhang:2010bm}
\begin{eqnarray}\label{Renorm-Lag}
\mathcal{L}_{\rm eff}+\delta\mathcal{L}_{\rm eff}
&=&-(\kappa^g,\kappa^\gamma)
\left(\begin{array}{ccc}1+\delta Z_{gg}&\delta Z_{g\gamma}\\
\delta Z_{\gamma g}&1+\delta Z_{\gamma\gamma}\\
\end{array}\right) \left(\begin{array}{c}O_g\\
O_\gamma\end{array}\right),
\end{eqnarray}
where the operators $O_i\ (i=g,\gamma)$ are defined as
$O_g=g_s\bar{q}\sigma^{\mu\nu}
T^a(f^g_{\mathrm{tq}}+ih^g_{\mathrm{tq}}\gamma_5)tG^a_{ \mu\nu}$,
$O_\gamma=e
\bar{q}\sigma^{\mu\nu}(f^{\gamma}_{\mathrm{tq}}+ih^\gamma_{
\mathrm{tq}}\gamma_5) tA_{\mu\nu}$, and $\delta Z_{gg}=\delta
Z_{\kappa^g_{\mathrm{tq}}/\Lambda}$, $\delta Z_{\gamma\gamma}=\delta
Z_{\kappa^\gamma_{\mathrm{tq}}/\Lambda}$. At the $\mathcal
{O}(\alpha_s)$ level, $\delta
Z_{\kappa^\gamma_{\mathrm{tq}}/\Lambda}$ is presented in
Eq.~(\ref{renkgamma}), and other renormalization constants are given
by
\begin{eqnarray}
\delta
Z_{\kappa^{g}_{tq}/\Lambda}&=&\frac{\alpha_s}{6\pi}
\Gamma(1+\epsilon)(4\pi)
^{\epsilon}\frac{1}{\epsilon_{UV}},\\
\delta Z_{g\gamma}&=&\frac{8\alpha_s}{9\pi}\Gamma(1+\epsilon)(4\pi)
^{\epsilon}\frac{1}{\epsilon_{UV}},\\
 \delta Z_{\gamma g}&=&0.
\end{eqnarray}

All the UV divergence are canceled exactly after the
renormalization. The remaining IR divergence of the virtual
corrections is the same as Eq.~(\ref{eq2}), except using the LO
amplitude including the contributions from both the electroweak and
the strong FCNC couplings instead of $M^B$.

For the real corrections, we still use the dipole subtraction method
as in above section. We don't repeat the detailed description here.
All the IR divergence from the virtual corrections are canceled
exactly. A criterion for isolated photon has been suggested in
Ref.~\cite{Frixione:1998jh}, which defines an IR-safe cross section
decoupled with hadronic fragmentation and at the same time allows
for complete cancelation of soft gluon divergence. For the case of
only one final-state massless parton, such criterion is equivalent
to the kinematic cut
\begin{eqnarray}
p^{j}_{T}<\frac{1-\cos\Delta R_{j\gamma}}{1-\cos\Delta
R_0}p^{\gamma}_{T},~~~~\textrm{for}\ \Delta R_{j\gamma}<\Delta R_0,
\end{eqnarray}
where $j$ stands for either the final-state (anti-)quark or the
final-state gluon. $\Delta R_{j\gamma}$ is the distance between the
parton and the photon in the rapidity-azimuthal angle plane. We
choose the cone-size parameters $\Delta R_0=0.7$ throughout our
calculation.

When we consider the mixing effects, the running of
$\kappa^{\gamma}$ is different from the one without mixing effects.
The running of $\kappa^{\gamma}$ and $\kappa^{g}$ are given by
~\cite{Zhang:2010bm}:
\begin{eqnarray}
\label{eq8} \kappa^g(\mu)&=&\kappa^g(\mu')
\eta^{\frac{2}{3\beta_0}},\\
\kappa^\gamma(\mu)&=&\kappa^\gamma(\mu')\eta^{\frac{4}{
3\beta_0}}
+\frac{16}{3}\kappa^g(\mu')\left(\eta^{\frac{4}{3\beta_0}}
-\eta^{\frac{2}{3\beta_0}}\right),
\end{eqnarray}
where $\eta=\alpha_s(\mu')/\alpha_s(\mu)$ and
$\beta_0=11-\frac{2}{3}n_f$ with $n_f=5$.

\section{Numerical Results}\label{s5}
\subsection{Process via the $tq\gamma$ FCNC
couplings without mixing effects}\label{ss2} Here we first consider
the top associated with $\gamma$ production via the $tq\gamma$ FCNC
couplings, including the NLO QCD effects on the total cross
sections, the scale dependence, and several important distributions
at both the Tevatron and LHC. For the numerical calculations of this
process, we take the SM parameters as
follow~\cite{Nakamura:2010zzi}:
\begin{equation}\label{para}
m_{t}=172.0{\rm GeV},\ \alpha_s(M_Z)=0.118,\ \alpha=1/128.921.
\end{equation}
And we set the electroweak FCNC couplings as:
\begin{equation}
\kappa^{\gamma}_{tu}/\Lambda=\kappa^{\gamma}_{tc}
/\Lambda=0.3{\rm
TeV}^{-1}.
\end{equation}
The running QCD coupling constant is evaluated at the three-loop
order~\cite{Nakamura:2010zzi} and the CTEQ6M PDF
set~\cite{Pumplin:2002vw} is used throughout the calculations of the
NLO (LO) cross sections. Unless specified, both the renormalization
and factorization scales are fixed to be the top quark mass.
Besides, we impose the photon transverse momentum cut $p_T>40{\rm
GeV}$ and pseudo-rapidity cut $|\eta|<2.5$. We have performed two
independent calculations for the virtual corrections and the
integrated dipole terms, and used the modified
MadDipole~\cite{Frederix:2008hu} package to generate the Fortran
code for the real corrections. The numerical results of the two
groups are in good agreement within the expected accuracy of our
numerical program.

\begin{table}[ht]
\begin{center}
\begin{tabular}{ccccc}
  \hline
  \hline
  FCNC coupling & $tu\gamma$  (LO) & $tu\gamma$  (NLO) &
$tc\gamma$  (LO) & $tc\gamma$  (NLO) \\
  \hline
  LHC $(\frac{\kappa/\Lambda}{{0.3\rm TeV}^{-1}})^2$ pb &
3.78 & 5.16 & 0.386 & 0.537 \\
  \hline
  Tevatron $(\frac{\kappa/\Lambda}{{0.3\rm TeV}^{-1}})^2$ fb
& 22.2  & 33.4  & 0.740 & 1.09 \\
  \hline
  \hline
\end{tabular}
\end{center}
\caption{The LO and NLO total cross sections for the single top
quark associated with $\gamma$ production via the $tq\gamma$ FCNC
couplings at both the LHC and Tevatron.} \label{t1}
\end{table}
In Table~\ref{t1}, we list some typical numerical results of the LO
and NLO total cross sections for the top quark associated with
$\gamma$ production via the electroweak FCNC couplings.

\begin{figure}[H]
  \subfigure{
    \begin{minipage}[b]{0.5\textwidth}
      \begin{center}
     \scalebox{0.8}{\includegraphics*[20,0][580,250]{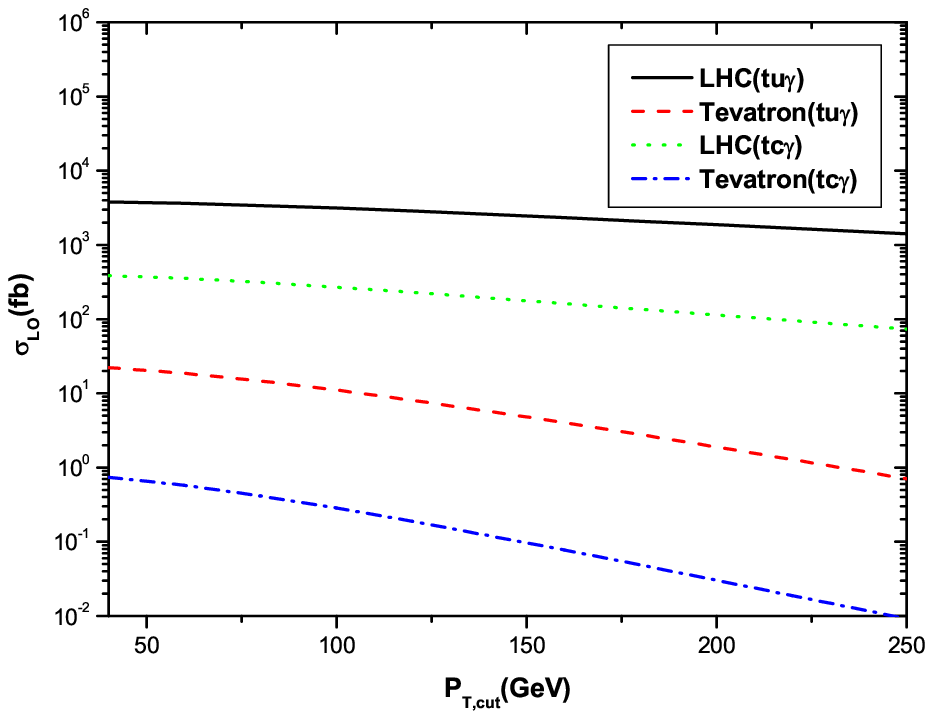}}
      \end{center}
    \end{minipage}}
  \subfigure{
    \begin{minipage}[b]{0.5\textwidth}
      \begin{center}
     \scalebox{0.8}{\includegraphics*[20,0][580,250]{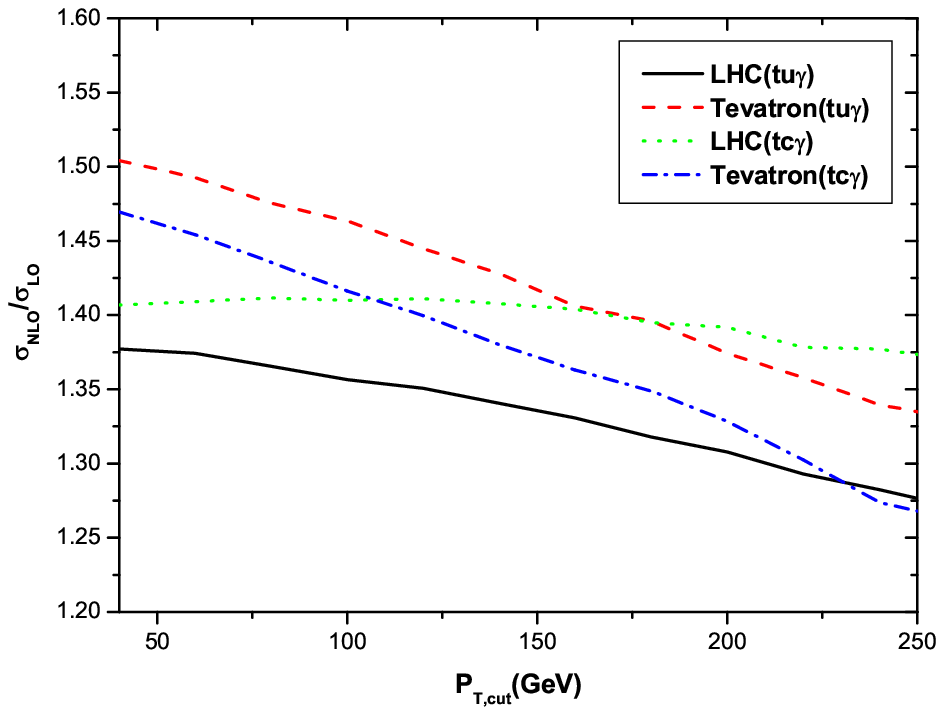}}
      \end{center}
    \end{minipage}}
    \caption{\label{ptcut}The LO total cross sections and NLO K factors
    as functions of the photon transverse momentum cut.}
\end{figure}

In Fig.~\ref{ptcut}, we show the LO total cross sections and the K
factors $\sigma_{NLO}/\sigma_{LO}$ as functions of the photon
transverse momentum cut, respectively. It can be seen that, for the
$tc\gamma$ coupling the NLO corrections can enhance the total cross
sections by about 50\% and 40\%, and for the $tu\gamma$ coupling by
about 50\% and 40\% at the Tevatron and LHC, respectively. And the K
factors decrease with the increasing transverse momentum cut.

\begin{figure}[H]
\begin{center}
\scalebox{1}{\includegraphics*[0,0][320,230]{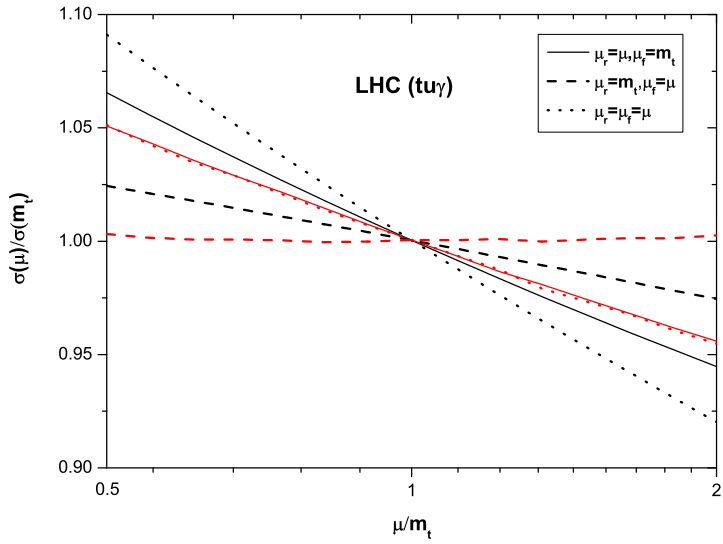}}
\caption[]{Scale dependence of the total cross sections at the LHC,
the black lines represent the LO results, while the red lines
represent the NLO results.} \label{scalel}
\end{center}
\end{figure}

\begin{figure}[H]
\begin{center}
\scalebox{1}{\includegraphics*[0,0][320,230]{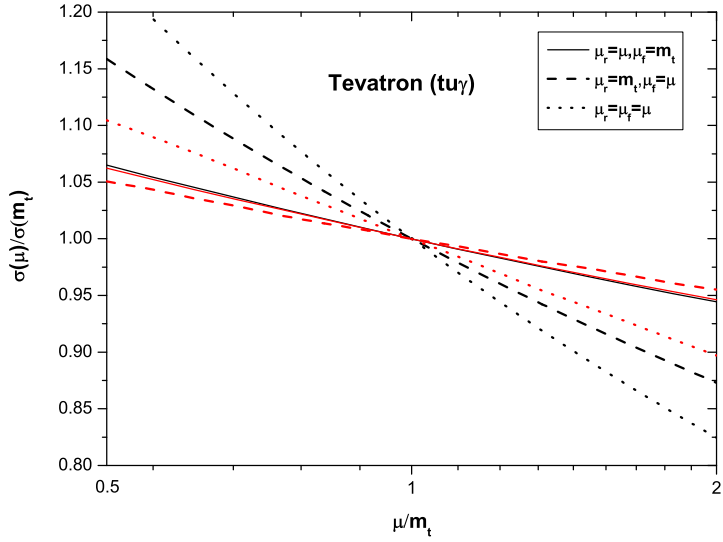}}
\caption[]{\label{scalet}Scale dependence of the total cross
sections at the Tevatron, the black lines represent the LO results,
while the red lines represent the NLO results.}
\end{center}
\end{figure}

In Fig.~\ref{scalel} and ~\ref{scalet} we present the scale
dependence of the LO and NLO total cross section for three cases:
(1) the renormalization scale dependence $\mu_r=\mu,\ \mu_f=m_t$,
(2) the factorization scale dependence $\mu_r=m_t,\ \mu_f=\mu$, and
(3) total scale dependence $\mu_r=\mu_f=\mu$. It can be seen that
the NLO corrections reduce the scale dependence significantly for
all three cases, which make the theoretical predictions more
reliable. For example, at the LHC for the $tu\gamma$ coupling, when
the scale $\mu$ varies from $0.5m_t$ to $2m_t$, the variations are
about $7\%$ and $5\%$ for case (1), $3\%$ and less than $1\%$ for
case (2), $9\%$ and $5\%$ for case (3), at the LO and NLO,
respectively.

\begin{figure}[H]
\begin{center}
\scalebox{1}{\includegraphics*[0,0][320,250]{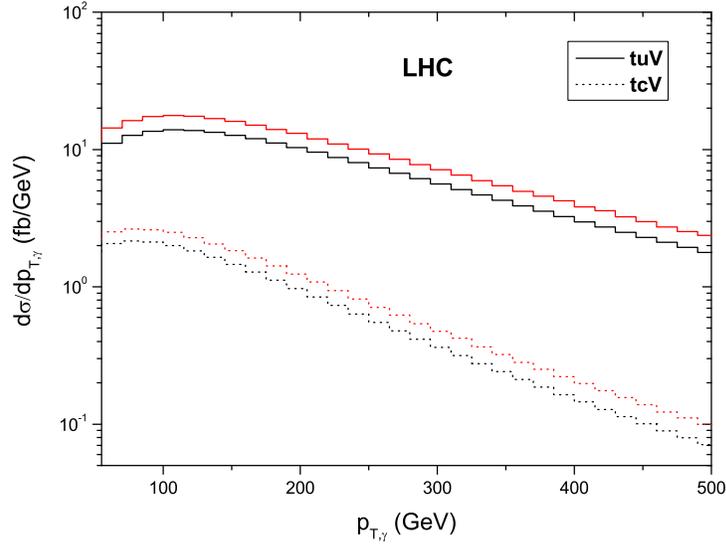}}
\caption[]{\label{ptdis}Transverse momentum distributions of the
photon, the black and red line represent the LO and NLO results of
the FCNC top quark associated $\gamma$ production, respectively.}
\end{center}
\end{figure}

\begin{figure}[ht]
\begin{center}
\scalebox{1}{\includegraphics*[0,0][320,230]{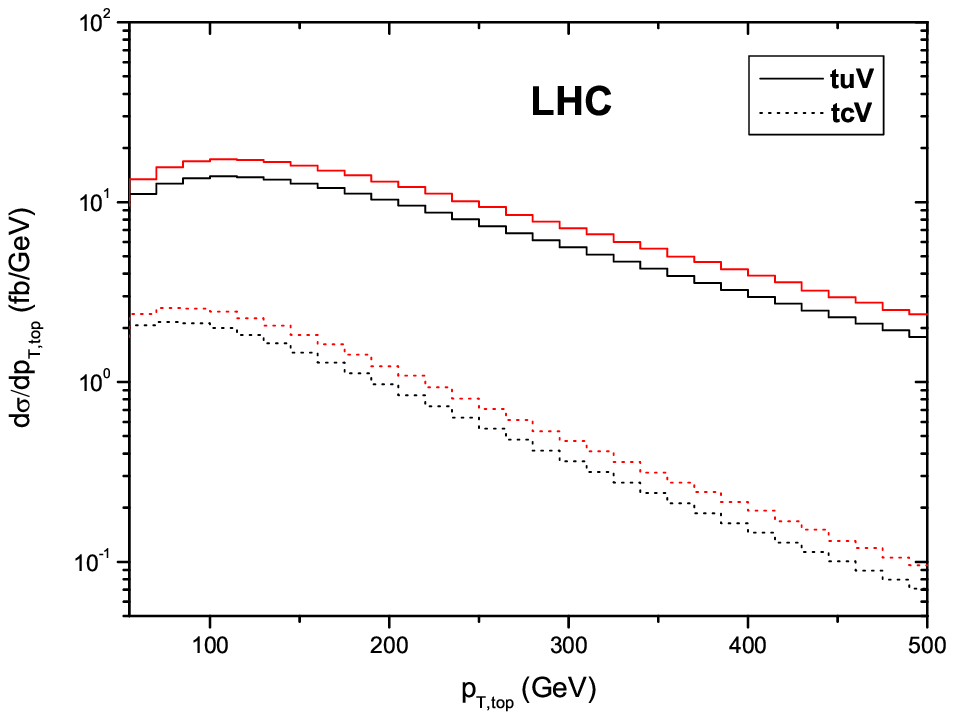}}
\caption[]{\label{toppt}Transverse momentum distributions of the top
quark, the black and red line represent the LO and NLO results of
the FCNC top quark associated $\gamma$ production, respectively.}
\end{center}
\end{figure}

\begin{figure}[H]
\begin{center}
\scalebox{1}{\includegraphics*[0,0][320,230]{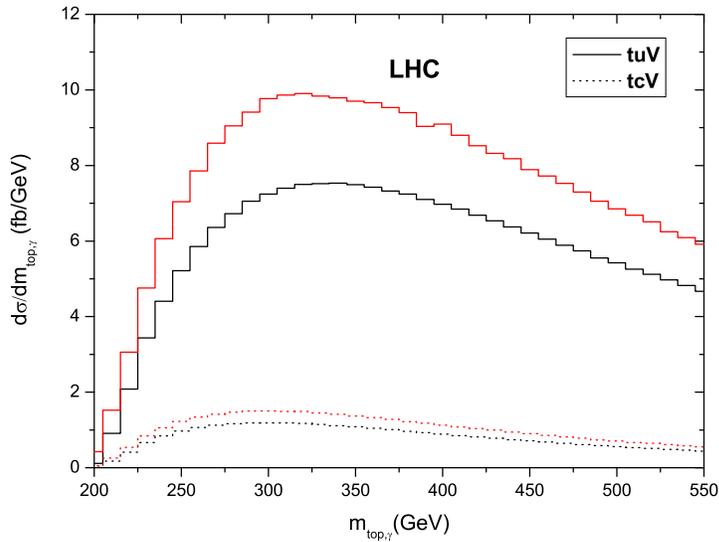}}
\caption[]{\label{invm}Invariant mass distributions of the photon
and the top quark, the black and red line represent the LO and NLO
results of the FCNC top quark associated $\gamma$ production,
respectively.}
\end{center}
\end{figure}

Fig.~\ref{ptdis} and~\ref{toppt} give the transverse momentum
distributions of the photon and the final state top quark,
respectively. We can see that the NLO corrections increase the
distributions of the FCNC top quark associated $\gamma$ production
in both high and low $p_T$ regions. Fig.~\ref{invm} shows the
invariant mass distributions of the photon and the top quark, where
there is a peak in the middle region in the invariant mass
distributions of this process. The NLO corrections do not change the
shapes of these distributions.

\subsection{Combination of the contributions from the $tq\gamma$ and $tqg$ FCNC couplings with mixing
effects}\label{ss1} In this subsection, we present the numerical
results of the single top associated with $\gamma$ production via
the electroweak and strong FCNC couplings, including the NLO QCD
effects and the mixing effects. For the numerical calculations, we
take the same SM parameters and kinematical cuts as above
subsection, and set the values of the FCNC couplings as follows:
\begin{eqnarray}
\kappa^{\gamma}_{tu}/\Lambda=\kappa^{\gamma}_{tc} /\Lambda=0.02{\rm
TeV}^{-1},\ \kappa^{g}_{tu}/\Lambda=\kappa^{g}_{tc}/\Lambda=0.01{\rm
TeV}^{-1}.
\end{eqnarray}

In Table~\ref{t2}, we show some typical numerical results of the LO
and NLO total cross sections for process induced by both the
electroweak and strong FCNC couplings with the mixing effects.
\begin{table}[ht]
\begin{center}
\begin{tabular}{ccccc}
  \hline
  \hline
  FCNC coupling & $tuV$  (LO) & $tuV$  (NLO) & $tcV$  (LO) &
$tcV$  (NLO) \\
  \hline
  LHC $(\frac{\kappa/\Lambda}{{0.01\rm TeV}^{-1}})^2$ fb &
27.8 & 42.7 & 3.13 & 5.61 \\
  \hline
  \hline
\end{tabular}
\end{center}
\caption{The LO and NLO total cross sections for process induced by
the $tq\gamma$ and $tqg$ FCNC couplings with mixing effects at the
LHC. Here
$f^{\gamma\ast}_{tq}f^{g}_{tq}+h^{\gamma\ast}_{tq}h^{g}_{tq} =1.$}
\label{t2}
\end{table}

From above results, we can see that the NLO effects are more
significant in the process induced by $tcV$ FCNC couplings than in
one induced by $tuV$ FCNC coupling. This is because that the
contributions from $gg\rightarrow \gamma t\bar{c}$ subprocess is
exactly the same as $gg\rightarrow \gamma t\bar{u}$ subprocess,
while the difference of the corresponding LO cross sections between
the above two processes is nearly 10 times. For example, at the LHC,
contributions from $gg\rightarrow \gamma t\bar{u}$ subprocess are
about $3\%$ of the corresponding LO cross section of $ug\rightarrow
\gamma t$, while the contributions from $gg\rightarrow \gamma
t\bar{c}$ subprocess are about $30\%$ of the corresponding LO cross
section of $cg\rightarrow \gamma t$.

\begin{figure}[ht]
\begin{center}
\scalebox{1}{\includegraphics*[0,0][320,230]{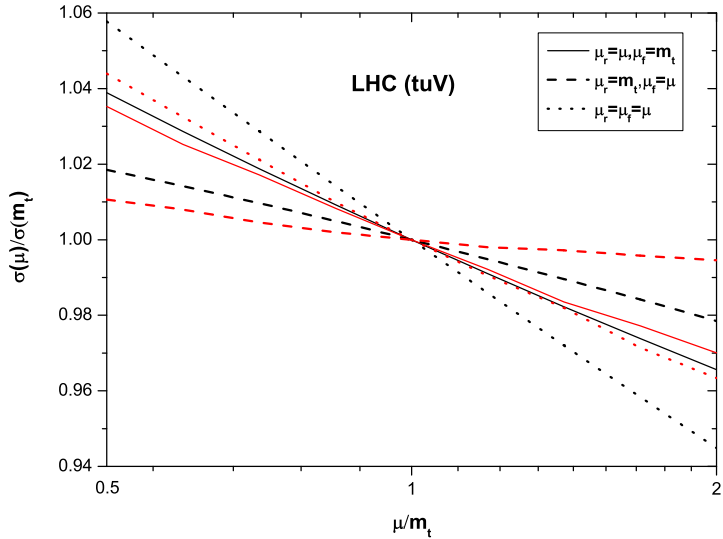}}
\caption[]{\label{scaleml}Scale dependence of the total cross
sections at the LHC, the black lines represent the LO results, while
the red lines represent the NLO results. Here
$\kappa^{\gamma}_{tu}=2\kappa^{g}_{tu}.$}
\end{center}
\end{figure}

Fig.~\ref{scaleml} shows the scale dependence of the total cross
sections for the top quark associated with $\gamma$ production via
the FCNC couplings with the mixing effects at the LHC. We can see
that the NLO QCD corrections reduce the dependence of the total
cross sections on the renormalization and factorization scale
significantly, as same as the case without the mixing effects.

\begin{figure}[H]
  \subfigure{
    \begin{minipage}[b]{0.5\textwidth}
      \begin{center}
\scalebox{0.8}{\includegraphics*[20,0][300,250]{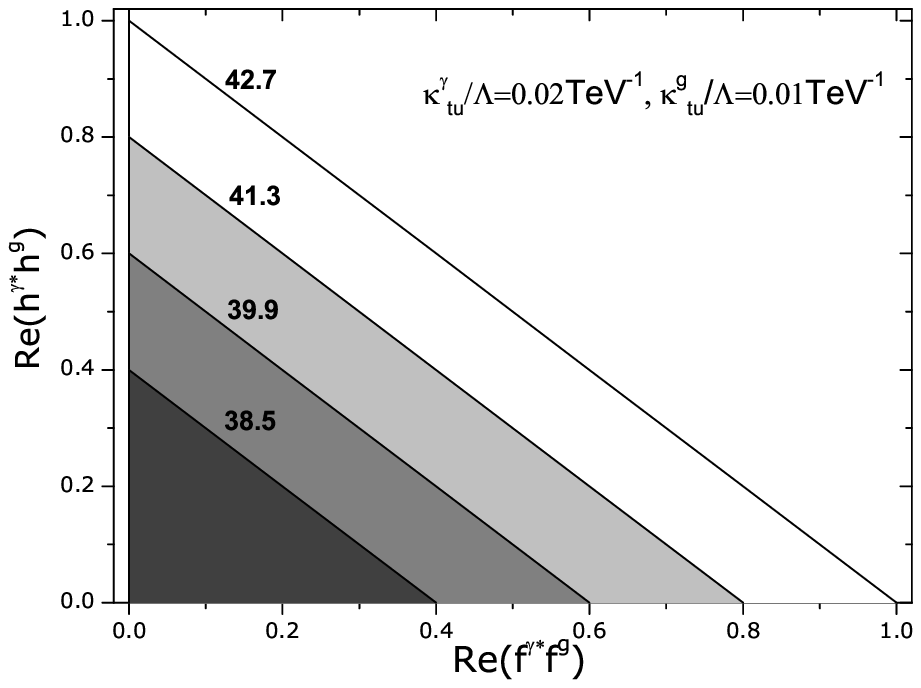}}
      \end{center}
    \end{minipage}}
  \subfigure{
    \begin{minipage}[b]{0.5\textwidth}
      \begin{center}
\scalebox{0.8}{\includegraphics*[20,0][300,250]{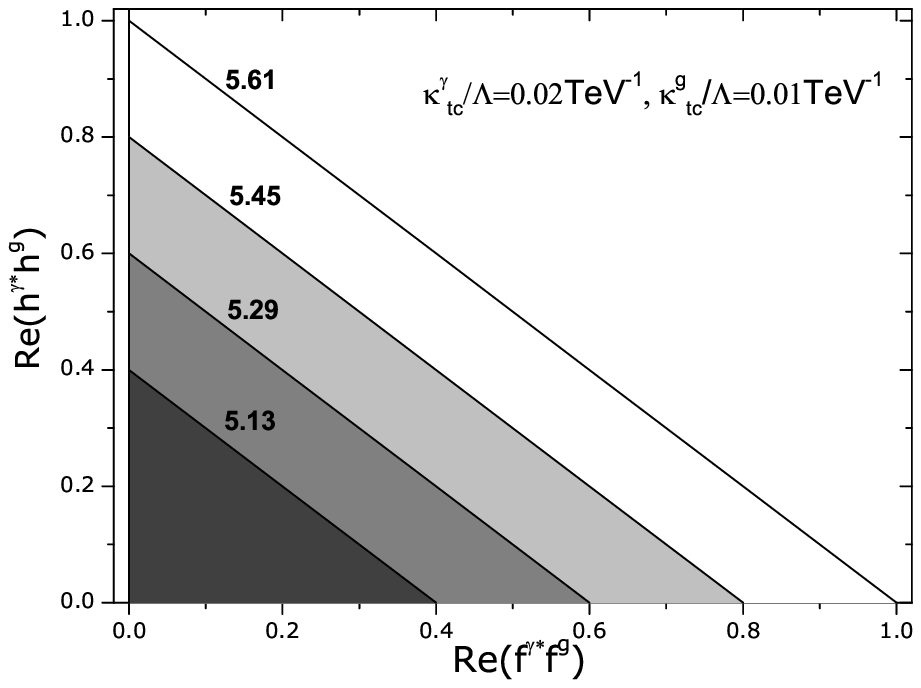}}
      \end{center}
    \end{minipage}}
    \caption{\label{contour}The contour curves of the total cross sections (fb) of the top
quark associated $\gamma$ production at the NLO including the mixing
effects versus the variables $\mathrm{Re}(f^{\gamma*}f^g)$ and
$\mathrm{Re}(h^{\gamma*}h^g)$. The left diagram represents the
process induced by $tuV$ couplings. And the right one shows the
process induced by $tcV$ couplings.}
\end{figure}

After considering the mixing effects, the total cross sections of
the top quark associated with $\gamma$ production via FCNC couplings
can be factorized as:
\begin{equation}
\sigma=A(\frac{\kappa^{\gamma}_{tq}}{\Lambda})^{2}+B(\frac{
\kappa^{g}_{tq}}{\Lambda})^{2}+
C(\frac{\kappa^{\gamma}_{tq}}{\Lambda})(\frac{\kappa^{g}_{tq
}}{\Lambda})\mathrm{Re}(f^{\gamma*}f^g+h^{\gamma*}h^g).
\end{equation}
where A, B and C represent the contributions from different
couplings and mixing effects. And, their numerical expressions at
the LHC can be written as
\begin{eqnarray}\label{ugc1}
\sigma^{tuV}_{LO}&=&\left[42.0(\frac{\kappa^{\gamma}_{tu}}{
\Lambda})^{2}+57.6(\frac{\kappa^{g}_{tu}}{\Lambda})^{2}+
26.4(\frac{\kappa^{\gamma}_{tu}}{\Lambda})(\frac{\kappa^{g}_
{tu}}{\Lambda})\mathrm{Re}(f^{\gamma*}f^g+h^{\gamma*} h^g)\right]
\mathrm{pb\cdot TeV^2},
\end{eqnarray}
\begin{eqnarray}\label{ugc2}
\sigma^{tuV}_{NLO}&=&\left[57(\frac{\kappa^{\gamma}_{tu}}{
\Lambda})^{2}+129(\frac{\kappa^{g}_{tu}}{\Lambda})^{2}+
35(\frac{\kappa^{\gamma}_{tu}}{\Lambda})(\frac{\kappa^{g}_{
tu}}{\Lambda})\mathrm{Re}(f^{\gamma*}f^g+h^{\gamma*} h^g)\right]
\mathrm{pb\cdot TeV^2};
\end{eqnarray}
\begin{eqnarray}\label{cgc1}
\sigma^{tcV}_{LO}&=&\left[4.3(\frac{\kappa^{\gamma}_{tc}}{
\Lambda})^{2}+9.5(\frac{\kappa^{g}_{tc}}{\Lambda})^{2}+
2.3(\frac{\kappa^{\gamma}_{tc}}{\Lambda})(\frac{\kappa^{g}_{
tc}}{\Lambda})\mathrm{Re}(f^{\gamma*}f^g+h^{\gamma*} h^g)\right]
\mathrm{pb\cdot TeV^2},
\end{eqnarray}
\begin{eqnarray}\label{cgc2}
\sigma^{tcV}_{NLO}&=&\left[6.0(\frac{\kappa^{\gamma}_{tc}}{
\Lambda})^{2}+24.1(\frac{\kappa^{g}_{tc}}{\Lambda})^{2}+
4.0(\frac{\kappa^{\gamma}_{tc}}{\Lambda})(\frac{\kappa^{g}_{
tc}}{\Lambda})\mathrm{Re}(f^{\gamma*}f^g+h^{\gamma*} h^g)\right]
\mathrm{pb\cdot TeV^2}.
\end{eqnarray}
In order to investigate the contributions from the mixing effects as
shown in Eqs.~(\ref{ugc1}),~(\ref{ugc2}),~(\ref{cgc1})
and~(\ref{cgc2}), we present the contour curves for the variables
$\mathrm{Re}(f^{\gamma*}f^g)$ and $\mathrm{Re}(h^{\gamma*}h^g)$, as
shown in Fig.~\ref{contour}. It can be seen that the total cross
sections at the NLO with the mixing effects increase slowly with
increasing of $\mathrm{Re}(f^{\gamma*}f^g)$ and
$\mathrm{Re}(h^{\gamma*}h^g)$.

\section{Conclusions}\label{s6}
We have calculated the NLO QCD corrections to the top quark
associated with $\gamma$ production via the $tq\gamma$ and $tqg$
FCNC couplings at hadron colliders, respectively, and we also
consider the mixing effects. Our results show that, the NLO QCD
corrections can enhance the total cross sections by about $50\%$ and
$40\%$ for the $tq\gamma$ couplings at the Tevatron and LHC,
respectively. If we combine the contributions from the $tq\gamma$,
$tqg$ FCNC couplings and the mixing effects, the NLO QCD corrections
can enhance the total cross sections by about $50\%$ for the
$tu\gamma$ and $tug$ FCNC couplings, and by about $80\%$ for the
$tc\gamma$ and $tcg$ FCNC couplings at the LHC. Moreover, the NLO
QCD corrections reduce the dependence of the total cross sections on
the renormalization or factorization scale significantly, which
leads to increased confidence in our theoretical predictions based
on these results. Besides, we also evaluate the NLO QCD corrections
to several important kinematic distributions, i.e., the transverse
momentum of the photon and the top quark, and the invariant mass of
the photon and the top quark, respectively. We find that the NLO
corrections are almost the same and do not change the shape of the
distributions.

\begin{acknowledgments}
This work was supported in part by the National Natural Science
Foundation of China, under Grants No. 10975004,  and No. 11021092.
\end{acknowledgments}

\end{document}